\RequirePackage{fix-cm}
\documentclass[smallcondensed]{svjour3}     
\usepackage{amsmath}
\usepackage{graphicx}
\begin{document}

\title{Gravitational lensing in the strong field limit for Kar's metric 
}



\author{Carlos A. Benavides \and \\
        Alejandro C\'ardenas-Avenda\~no \and \\
        Alexis Larranaga 
}


\institute{Carlos Benavides \at
              Universidad Nacional de Colombia\\
              Av. Carrera 30 \# 45-03. C\'od. postal 111321 \\
              Tel.: (+57 1) 316 5000\\
              \email{cabenavidesg@unal.edu.co}           
           \and
            Alejandro C\'ardenas-Avenda\~no\at
            Fundacion Universitaria Konrad Lorenz\\
              Carrera 9 Bis No. 62 - 43\\
              Tel.:(+57 1) 347 23 11\\
              \email{alejandro.cardenasa@konradlorenz.edu.co}
           \and
            Alexis Larranaga \at
            Universidad Nacional de Colombia\\
            Av. Carrera 30 \# 45-03. C\'od. postal 111321 \\
              Tel.: (+57 1) 316 5000\\
              \email{ealarranaga@unal.edu.co}           
}

\date{Received: date / Accepted: date}

\maketitle

\begin{abstract}
In this paper we calculate the strong field limit deflection angle
for a light ray passing near a scalar charged spherically symmetric
object, described by a metric which comes from the low-energy limit
of heterotic string theory. Then, we compare the expansion parameters
of our results with those obtained in the Einstein's canonical frame,
obtained by a conformal transformation, and we show that, at least
at first order, the results do not agree.
 
\end{abstract}

\section{Introduction}
\label{intro}

The General Theory of Relativity (GR) is the best theory of gravitation available. It establishes a new conception of 
space and time and describes how curvature acts on matter, to manifest itself as gravity, and how energy and momentum 
influence spacetime to create curvature: \textit{space tells matter how to move; matter tells spacer how to curve} 
\cite{Gravitation}. It has also passed several experimental tests in the weak 
field limit; however, it has not been yet tested in the strong gravitational field regime \cite{will2006,Psaltis2009}.

GR has important astrophysical implications among which the deflection of 
light is one of the most important \cite{Dover}. Light deflection in weak gravitational field is well-known since 1919, 
when Eddington observed the light deflection by the sun \cite{Dover}, in lensing of quasars by foreground galaxies 
\cite{Walsh1979}, in the formation of giant arcs in galaxy clusters \cite{Fort1994} and more recently in galactic 
microlensing \cite{Narasimha1995}. Now it is an important phenomenon in the panorama of astronomical observations 
\cite{Bozza2002}. 

The first study regarding light deflection in the strong regime was made by Darwin in Refs. \cite{Darwin,Darwin2} 
where he initiated a theoretical research on gravitational lensing resulting from large deflection of light in the 
vicinity of the photon sphere of Schwarzschild spacetime. These studies were extended to a general static spherically 
symmetric spacetime 
by Atkinson in Ref. \cite{Atkiston}. Nevertheless, the subject of strong gravitational field lensing remained in a 
quiet stage for two important reasons: 1) Darwin's calculations showed that the images are very difficult to be 
observed with the available observational facilities at that time, 2) the known gravitational lens equation was not 
adequate for the 
study of lensing due to large deflection of light \cite{Virbhadra2009}. However, as astronomical techniques are 
improving fast, the possibility of testing the nature of astrophysical black hole candidates with current and future 
observations has recently become an active research field, since in the following years the technique of very long 
baseline interferometry (VLBI) and, in a longer term, the (sub)millimeter VLBI \textquotedblleft Event Horizon 
Telescope\textquotedblright{} will produce images of the Galactic center emission capable to see the silhouette 
predicted by general relativistic lensing \cite{Deoleman2009,Fish2009}. Therefore, testing the gravitational field in 
the vicinity of a compact massive object, such as a black hole or a neutron star, could be a possible avenue for such 
investigations. In this sense, the importance of gravitational lensing in strong fields is highlighted by the 
possibility of testing the full GR in a regime where the differences with non-standard
theories would be manifest, helping the discriminations among the
various theories of gravitation \cite{will2008,Broderick2014,Capozziello}.
For this reason, the scientific community has been interested in the
lensing properties near the photon sphere i.e. strong field limit.

The strong field limit lensing regime was first defined in Ref. \cite{Virbhadra2000}, where the authors studied 
the strong gravitational lensing due to a Schwarzschild black hole showing that, apart from the primary and 
the secondary images, there exists a sequence of images on both sides of the optic axis; named \textit{relativistic 
images}. One of the most important contributions of that 
paper was a lens equation that allows, for a large bending of light near a black hole, to model the 
Galactic supermassive ‘‘black hole’’ as a Schwarzschild lens and to study point source lensing in the strong 
gravitational 
field region \cite{Virbhadra2000}. Later on, Virbhadra and Ellis in 2002 modeled massive dark objects in the galactic 
nuclei as spherically symmetric static naked singularities in the Einstein Massless Scalar (EMS) field theory 
\cite{Virbhadra2002}. In the same year Bozza, in Ref \cite{Bozza2002}, provided a general method to 
extend the strong field limit to a generic static spherically symmetric spacetime inspired by the previous works 
\cite{Capozziello,Virbhadra2002}. Bozza expanded the deflection angle near the photon sphere and showed 
that the divergence is logarithmic for all spherically symmetric metrics. According to the author, this method can be 
applied to any spacetime in any theory of gravitation, as long as the photons satisfy the geodesic equation 
\cite{Bozza2002}. Using 
this method it is possible to discriminate among different types of black holes on the grounds of their strong field 
gravitational lensing properties e.g., studying the properties of the relativistic images it may be possible to 
investigate the regions immediately outside of the event horizon because the parameters of the strong field limit 
expansion are directly connected with observables \cite{Bozza2002}. For this reason, the  strong field limit has been used to 
estimate the deflection angle and the position of relativistic images produced by different types of black holes 
(see for instance, Refs. \cite{Varios} and references therein, where gravitational 
lensing is not conceived as a weak field phenomenon).

In this paper, using the formalism presented in Refs. \cite{Virbhadra2002,Bozza2002}, we calculate the 
strong field 
limit deflection angle
for a light ray passing near a scalar charged spherically symmetric
object, described by the metric proposed by Kar in Ref. \cite{Kar1999}, which comes from the low-energy limit of the
heterotic string theory in the Jordan frame. The metric used in this paper is 
considered 
equivalent to the Janis-Newman-Winicour (JNW) metric\footnote{This metric was 
obtained independently by Wyman in 1981 in
Ref. \cite{Wyman}, but its equivalence  with the JNW's metric was not known until 1997 by Virbhadra in 
Ref. \cite{Virbhadra1997}. See Refs. \cite{VirbhadraNaked} for further details regarding the gravitational lensing in this metric and classification of its singularities.} \cite{JNW1968} under a conformal 
transformation, since it is possible to rewrite it in the Einstein canonical frame (as written in Ref. 
\cite{Virbhadra1997}) by employing the standard relations
between the two metrics 
$g_{\mu\nu}^{str}=e^{2\phi}g_{\mu\nu}^{E}$
\cite{Kar1999}. This result allows us to compare the frames from the strong field limit point of view, since Bozza in 
Ref. \cite{Bozza2002} calculated the same angle for the JNW's metric and obtained different results. At least at first 
order the difference agrees with the ideas presented by Alvarez and Conde in Ref. \cite{AlvarezConde2002}, where they 
argued that the equivalence of the frames,
for the description of the gravitational effects, is only obtained
when all functions involved are smooth, condition which is not satisfied
by the method used here, since the deflection angle in the strong field limit
diverges around the photon sphere. 

This paper is organized as follows. In section \ref{sec:Strong-field-expansion} we review briefly the method used, 
write down the relevant equations and then apply the strong field limit expansion to Kar's metric and in section
\ref{sec:Strong-field-expansionJNW} for the JNW's metric. In section 
\ref{sec:Analysis} we compare the results obtained in the Jordan frame with those obtained
by Bozza in the Einstein frame and we calculate, as an example, the tangential magnification for the case 
of the Galactic supermassive ''black hole''. Finally, section \ref{sec:Discussion} contains the discussion of the 
results. The technical part of the paper is presented as an Appendix \ref{sec:Appendix}, which has a detailed 
description of the calculations to obtain the deflection angle for Kar's metric in the strong field limit.

\section{\label{sec:Strong-field-expansion}Strong Field Expansion for Kar's
Metric}
A general spherical symmetric spacetime is described by the line element \cite{Virbhadra1998} 

\begin{equation}
ds^{2}=A(x)dt^{2}-B(x)dx^{2}-C(x)(d\theta^{2}+\sin^{2}\theta d\phi^{2}).\label{sphericallmetric}
\end{equation}
In order to obtain the deflection of a light beam it is necessary
to consider the motion of a freely falling photon in a static isotropic
gravitational field. From the geodesic equation and the line element
(\ref{sphericallmetric}) it is possible to obtain, as a function
of the closest distance of approach $x_{0}$, the following expression for the
deflection angle \cite{Virbhadra1998} 

\begin{equation}
\hat{\alpha}(x_{0})=2\int_{x_{0}}^{\infty}\sqrt{\frac{B(x)}{C(x)}}\left[\frac{C(x)}{C(x_{0})}\frac{A(x_{0})}{A(x)}-1\right]^{-\frac{1}{2}}dx-\pi.\label{deflexion angle}
\end{equation}

According to equation (\ref{deflexion angle}), a photon coming from
infinity with some impact parameter $u=\sqrt{\frac{C(x_{0})}{A(x_{0})}}$ \cite{Virbhadra1998}
will be deviated when it is approaching the black hole, it will reach
$x_{0}$ and then emerge in another direction. By decreasing $x_{0}$
the deflection angle increases until it exceeds $2\pi$, when the
photon gives a complete loop around the black hole. By decreasing
the impact parameter $u$ the photon will wind several times before
emerging. Finally, for $x_{0}$ equal to the photon's sphere radius ($x_m$), the deflection angle diverges and the photon is captured.

The formalism presented in Refs. \cite{Virbhadra2002,Bozza2002}  is used to calculate the deflection angle in the 
strong
field limit for a generic static spherically and symmetric spacetime taking the photon
sphere as the starting point. The photon sphere is the region of spacetime
where gravity is strong enough that photons are forced to travel in
orbits \cite{Virbhadra2002}. This means that Einstein's bending angle of a light ray with
the closest distance of approach $x_{0}$ becomes unboundedly large
as it tends to $x_{m}$. In this sense, the method requires that
the photon sphere equation \cite{Claudel}
\begin{equation}
C'(x)A(x)=A'(x)C(x)\label{eq:photon sphere}
\end{equation}
has at least one positive solution. Here a prime represents the derivative with respect to $x$. In general, equation 
(\ref{eq:photon sphere}) has several solutions; however, we will take the largest root as the radius of the photon 
sphere and denote it by $x_{m}$, as it is defined in \cite{Bozza2002}. To obtain the deflection angle in the strong 
field limit, the equation (\ref{deflexion angle}) is written in the following convenient form \footnote{By making 
$y=A(x)$ and 
$z=\frac{y-y_0}{1-y_0}$.} 

\begin{equation}
\hat{\alpha}(x_{0})=\int_{0}^{1}R(z,x_{0})f(z,x_{0})dz-\pi=I(x_{0})-\pi\label{newalpha}
\end{equation}
where,
\begin{equation}
R(z,x_{0})=\frac{2\sqrt{By}}{CA'}(1-y_{0})\sqrt{C_{0}}\label{R(z,x)},
\end{equation}

\begin{equation}
f(z,x_{0})=\frac{1}{\sqrt{y_{0}-[(1-y_{0})z+y_{0}]\frac{C_{0}}{C}}}.\label{f(x,z)}
\end{equation}

All functions with the subscript $0$ are evaluated at $x=x_{0}$ and without it are evaluated at $x=A^{-1}[(1-y_{0})z+y_{0}]$. The prime $'$ is the derivative with respect to $x$. Equation (\ref{deflexion angle}) expressed in this form makes possible to identify  the function which contains the divergence. 
According to Eqns. (\ref{R(z,x)}) and (\ref{f(x,z)}), the function $R(z,x_{0})$ is regular for values of $z$ and $x_{0}$. However, the function $f(z,x_{0})$ diverges for $z\to0$. In this sense, to obtain the order of divergence of the integrand it is necessary to expand the argument of the square root in $f(z,x_{0})$ to the second order in $z$. Therefore, for $z\to0$ the function $f(z,x_{0})$ can be approximate to 

\begin{equation}
f_{0}(z,x_{0})=\frac{1}{\sqrt{\alpha z+\beta z^{2}}}\label{f0},
\end{equation}
where $\alpha$ and $\beta$ are expressed by 

\begin{equation}
\alpha=\frac{1-y_{0}}{C_{0}A'_{0}}(C'_{0}y_{0}-C_{0}A'_{0})\label{alpha_beta}
\end{equation}
and
\begin{equation}
\beta=\frac{(1-y_{0})^{2}}{2C_{0}^{2}A_{0}^{'3}}[2C_{0}C'_{0}A_{0}^{'2}+(C_{0}C''_{0}-2C_{0}^{'2})y_{0}A'_{0}-C_{0}C'_{0}y_{0}A''_{0}].\label{beta en terminos de la metrica}
\end{equation}

When $\alpha$ is not zero, the leading order of the divergence in $f_{0}$ is $z^{-\frac{1}{2}}$, which can be integrate to give a finite result. When $\alpha$ vanishes, the divergence is $z^{-1}$, which makes the integral diverge \cite{Bozza2002}. If we examine
the form $\alpha$, we see that it vanish at $x_{0}=x_{m}$. Each photon having $x_{0}<x_{m}$ is captured by the central object and can not emerge back.
 
Having found the function which contains the divergence, the next step is to split $I(x_{0})$
into two pieces: one part containing the divergence, $I_{D}(x_{0})$,
and the other being the original integral with the divergence subtracted,
$I_{R}(x_{0})$
\begin{eqnarray}
I(x_{0}) & = & \int_{0}^{1}R(z,x_{0})f(z,x_{0})dz=I_{D}(x_{0})+I_{R}(x_{0}).\label{ID_IR}
\end{eqnarray}  
Using this idea, in Ref. \cite{Bozza2002} was shown that the divergence of equation (\ref{deflexion angle}) is 
logarithmic for all spherically symmetric metrics and has the form,

\begin{equation}
\alpha(\theta)=-\overline{a}\ln\left[\frac{\theta D_{OL}}{u_{m}}-1\right]+\overline{b},\label{deflection_as_a 
function_of_theta}
\end{equation}
where
\begin{eqnarray}
\bar{a} & = & \frac{R(0,x_{m})}{2\sqrt{\beta_{m}}},\nonumber \\
\bar{b} & = & b_{R}+\bar{a}\ln\left[\frac{2\beta_{m}}{y_{m}}\right]-\pi
\label{a_and_bdefletion theta}
\end{eqnarray}
and
\begin{equation}
u_m=\sqrt{\frac{C(x_{m})}{A(x_{m})}}\label{u_m}.
\end{equation}
As we can see $\bar{a}$, $\bar{b}$, $u_m$ depend on the metric functions
evaluated at $x_m$, $D_{OL}$ is the distance between the lens and
the observer and
\begin{equation}
b_{R}=I_{R}(x_{m}).\label{b_R}
\end{equation}

By the procedure described above we calculated the deflection angle in the strong field limit for the metric proposed 
by Kar in Ref. \cite{Kar1999}, which has the form
\begin{equation}
ds_{str}^{2}=\left(1-\frac{2\eta}{r}\right)^{\frac{m+\sigma}{\eta}}dt^{2}-\left(1-\frac{2\eta}{r}\right)^{\frac{
(\sigma-m)}{\eta}}dr^{2}-\left(1-\frac{2\eta}{r}\right)^{1+\frac{\sigma-m}{\eta}}r^{2}d\Omega^{2},\label{karmetric}
\end{equation}
where $m$ is the mass, $\sigma$ is the scalar charge and $\eta$
is given by $\eta^{2}=m^{2}+\sigma^{2}$. For $\sigma=0$ this solution
reduces to the Schwarzschild solution. Using geometrized units (the
gravitational constant $G=1$ and the speed of light in vacuum $c=1$)
and introducing a radial distance defined as $x=\frac{r}{2\eta}$ and
$x_{o}=\frac{r_{0}}{2\eta}$, equation (\ref{karmetric}) takes the
form 

\begin{equation}
ds_{str}^{2}=\left(1-\frac{1}{x}\right)^{\frac{m+\sigma}{\eta}}dt^{2}-\left(1-\frac{1}{x}\right)^{\frac{(\sigma-m)}{\eta}}dx^{2}-\left(1-\frac{1}{x}\right)^{1+\frac{\sigma-m}{\eta}}x^{2}d\Omega^{2}.\label{kar_metric_I}
\end{equation}

Nevertheless, in order to discuss gravitational lensing in the strong field limit
for equation (\ref{kar_metric_I}), it is more convenient to express it
in terms of a single parameter. Using the relation $\eta^{2}=m^{2}+\sigma^{2}$,
we defined $\zeta=\frac{\sigma}{\eta}$ and $\gamma=\frac{m}{\eta}$
(which is the JNW's parameter used Ref. \cite{Bozza2002}) so that $\gamma^{2}+\zeta^{2}=1$. If we choose $\zeta$ as the 
parameter, the Kar's metric can be expressed as

{\scriptsize{}
\begin{equation}
ds_{str}^{2}=\left(1-\frac{1}{x}\right)^{\zeta+\sqrt{1-\zeta^{2}}}dt^{2}-\left(1-\frac{1}{x}\right)^{\zeta-\sqrt{1-\zeta^{2}}}dx^{2}-\left(1-\frac{1}{x}\right)^{1+\zeta-\sqrt{1-\zeta^{2}}}x^{2}d\Omega^{2},\label{Kar metric II}
\end{equation}
}{\scriptsize}
were $\gamma=\sqrt{1-\zeta^{2}}$. Note that for $\zeta=0$ the metric (\ref{Kar metric II})
reduces to Schwarzschild. Therefore, in Kar's metric (\ref{Kar metric II})  the functions
for a spherically symmetric metric are 

\begin{equation}
A(x)=\left(1-\frac{1}{x}\right)^{\zeta+\sqrt{1-\zeta^{2}}},
\end{equation}

\begin{equation}
B(x)=\left(1-\frac{1}{x}\right)^{\zeta-\sqrt{1-\zeta^{2}}}
\end{equation}
and
\begin{equation}
C(x)=\left(1-\frac{1}{x}\right)^{1+\zeta-\sqrt{1-\zeta^{2}}}x^{2}.
\end{equation}

From equation (\ref{alpha_beta}) with $\alpha(x_m)=0$, the radius of the photon sphere as a function of $\zeta$ 
is 

\begin{equation}
x_{m}=\sqrt{1-\zeta^{2}}+\frac{1}{2}.\label{photon sphere kar}
\end{equation}

For $\zeta=0$ equation (\ref{photon sphere kar}) reduces to $x_{m}=\frac{3}{2}$,
which is the value of the Schwarzschild's photon sphere radius \cite{Virbhadra2000}. 

From Eqns. (\ref{photon sphere kar}) and (\ref{alpha_beta})
we obtain that

{\scriptsize{}
\[
R(0,x_{m})=\frac{2\sqrt{1-\zeta^{2}}+1}{\sqrt{1-\zeta^{2}}+\zeta}\left\{ 
\frac{\left(\frac{2\sqrt{1-\zeta^{2}}-1}{2\sqrt{1-\zeta^{2}}+1}\right)^{\frac{1+\zeta-\sqrt{1-\zeta^{2}}}{2}}
-\left(\frac{2\sqrt{1-\zeta^{2}}-1}{2\sqrt{1-\zeta^{2}}+1}\right)^{\frac{1+3\zeta-\sqrt{1-\zeta^{2}}}{2}}}{y_{m}^{\frac{
\zeta}{\sqrt{1-\zeta^{2}}+\zeta}}}\right\},
\]
}{\scriptsize}
where 
\[
y_{m}=\left(\frac{2\sqrt{1-\zeta^{2}}-1}{2\sqrt{1-\zeta^{2}}+1}\right)^{\sqrt{1-\zeta^{2}}+\zeta}
\]

and 

\begin{equation}
\beta_{m}=\frac{1}{4}\frac{\left[\left(2\sqrt{1-\zeta^{2}}+1\right)^{\zeta+\sqrt{1-\zeta^{2}}}-\left(2\sqrt{1-\zeta^{2}}-1\right)^{\zeta+\sqrt{1-\zeta^{2}}}\right]^{2}}{(\sqrt{1-\zeta^{2}}+\zeta)^{2}(3-4\zeta^{2})^{\sqrt{1-\zeta^{2}}+\zeta-1}}.\label{beta_m_S_kar}
\end{equation}
Thus, from equation (\ref{a_and_bdefletion theta}) we obtain that
$\bar{a}=1$, which is the same value obtained for JNW's metric \cite{Bozza2002}.

In order to obtain $b_{R}$ we expand equation (\ref{b_R}) up to
first order in $\zeta$, around $\zeta=0$, obtaining that

\begin{equation}
b_{R}=2\ln6(2-\sqrt{3})-0.226\zeta\label{b_R para Kar}
\end{equation}
and using equation (\ref{a_and_bdefletion theta})
\begin{equation}
\bar{b}=2\ln(6(2-\sqrt{3}))-0.226\zeta-\pi+\ln\left[\frac{2\beta_{m}}{y_{m}}\right].
\end{equation}
Making $x_0=x_m$ in $u=\sqrt{\frac{C(x_{0})}{A(x_{0})}}$, we obtain that the impact parameter is 

\[
u_{m}=\frac{1}{2}\frac{(2\sqrt{1-\zeta^{2}}-1)^{\frac{1}{2}-\sqrt{1-\zeta^{2}}}}{(2\sqrt{1-\zeta^{2}}+1)^{-\frac{1}{2}-\sqrt{1-\zeta^{2}}}}.
\]
Finally, using equation (\ref{deflection_as_a function_of_theta}),
the deflection angle in the strong field limit for Kar's metric in
terms of the parameter $\zeta$ is 

\begin{equation}
\hat{\alpha}=-\ln\left[\frac{u}{u_{m}}-1\right]+2\ln(6(2-\sqrt{3}))-0.226\zeta-\pi+\ln\left[\frac{2\beta_{m}}{y_{m}}\right],\label{kar deflection angle}
\end{equation}
where $u$ is the impact parameter. The last equation can be written in terms of the angular 
separation of the image $\theta$ by the relation $\theta=\frac{u}{D_{OL}}$, where $D_{OL}$ is the 
distances between the lens and the observer (see Eqn. (\ref{alphaappend}) for details).

\section{\label{sec:Strong-field-expansionJNW}Strong Field Expansion for JNW's
Metric}

The JNW's metric \cite{Virbhadra1997}
\begin{equation}
ds^2=\left(1-\frac{1}{x}\right)^\gamma 
dt^2-\left(1-\frac{1}{x}\right)^{-\gamma}dx^2-\left(1-\frac{1}{x}\right)^{1-\gamma},\label{JNW}
\end{equation}   

has a photon sphere, $x_{m}$, which now expressed as a function of 
$\gamma$ is \cite{Bozza2002}
\begin{equation}
x_{m}=\frac{2\gamma+1}{2},\label{JNWxm}
\end{equation}
which implies that the photon sphere exists only for $\frac{1}{2}<\gamma\leq 1$ \cite{Claudel}. 
For this interval, a photon coming from infinity is deflected through an unboundedly large angle; this means, that the 
photon passes many times around the singularity as the closest distance of approach tends to $x_m$. 

The deflection angle in the strong field limit obtained for JNW's metric in Ref. \cite{Bozza2002} is 
\begin{equation}
\widehat{\alpha}(\theta)=\overline{a}\ln\left(\frac{\theta D_{OL}}{u_m}-1\right)+\overline{b},
\end{equation}
where $\overline{a}=1$ and 
{\scriptsize{}
\begin{equation}
\overline{b}=2\ln[6(2-\sqrt{3})]-0.1199(\gamma-1)-\pi+2\ln\left(\frac{(2\gamma+1)[(2\gamma+1)^\gamma-(2\gamma-1)^\gamma]
^2}{2\gamma^2(2\gamma-1)^{2\gamma-1}}\right).\label{BozzaI}
\end{equation}
}{\scriptsize}
However, note that the values of $\beta_m$ and, in consequence, the value of $\overline{b}$ (see equation 
(\ref{a_and_bdefletion theta})) differs from Eqns. (72) and (75) of Ref.  \cite{Bozza2002}. Our calculations show 
that

\begin{equation}
\beta_m=\frac{[(2\gamma+1)^\gamma-(2\gamma-1)^\gamma]^2}{4\gamma^2(4\gamma^2-1)^{\gamma-1}}\label{correction}
\end{equation}
thus,
{\scriptsize{}
\begin{equation}
\overline{b}=2\ln[6(2-\sqrt{3})]-0.1199(\gamma-1)-\pi+\ln\left(\frac{(2\gamma+1)[(2\gamma+1)^\gamma-(2\gamma-1)^\gamma]
^2,}{2\gamma^2(2\gamma-1)^{2\gamma-1}}\right)\label{correctionI}
\end{equation} 
}{\scriptsize}
which differs from a factor of two in the logarithmic. For $\gamma=1$, Eqns. (\ref{correction}) and 
(\ref{correctionI}) reduce to those of Schwarzschild \cite{Bozza2002}.

\begin{figure}[ht]
\begin{centering}
\includegraphics[scale=0.55]{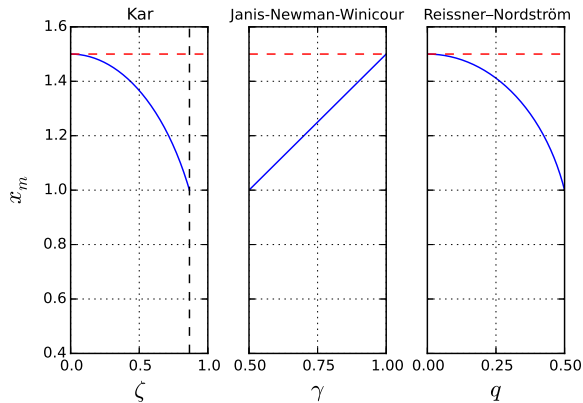}
\par\end{centering}

\protect\caption{\label{fig:Behavior-of-the} Behavior of the photon sphere, $x_{m}$,
for Kar's, JNW's and RN's metrics. In all figures, the red dashed horizontal lines
are the Schwarzschild limit.}
\end{figure}

 
In order to discuss gravitational lensing in the strong field limit
for Kar's metric (\ref{Kar metric II}) and compared it with (\ref{JNW}), it is necessary to find the values of 
$\zeta$ where equation (\ref{photon sphere kar}) has a
solution. Using the analysis made in \cite{Claudel} we found that the photon sphere equation has solution only for 
$0\leq\zeta<\frac{\sqrt{3}}{2}$, i.e., for $0\leq\sigma^{2}<3m^{2}$. It is easy to obtain the same interval for $\zeta$ 
using $\frac{1}{2}<\gamma\leq1$ (the interval for JNW's metric) and recalling that $\gamma=\sqrt{1-\zeta^{2}}$. The 
behaviors of 
the photon sphere for Kar, JNW and Reissner$-$Nordström (RN), also presented in Ref. \cite{Bozza2002}, metrics 
as a function of $\zeta$, $\gamma$ and $q$ (which is the charge in the RN metric), respectively, are plotted in Fig. 
\ref{fig:Behavior-of-the}. 

In Fig. \ref{fig:Behavior-of-,} is plotted the values of $u_{m}$, $\bar{a}$ and $\bar{b}$ in terms of $\zeta$ for Kar's 
metric. From 
the figure we see that $\bar{a}=1$ is constant and has the same value of that of JNW's metric \cite{Bozza2002}. Note 
that 
$\overline{b}$ has a minimum value at $z=0$, increases as $\zeta$ increases, and for $\zeta=\sqrt{3}/2$ the value of 
$\overline{b}$ diverges. In the same figure, the impact parameter $u_m$ decreases as $\zeta$ increases until the value 
$\zeta=\sqrt{3}/2$ is reached, where $u_m$ diverges. Finally, it is 
clear that each parameter reduces to those of Schwarzschild for $\zeta=0$.

\begin{figure}[ht]
\begin{centering}
\includegraphics[scale=0.5]{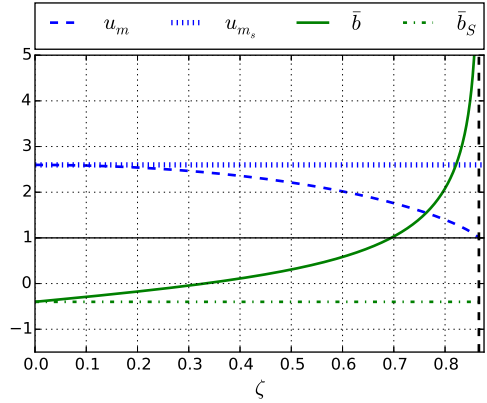}
\par\end{centering}

\protect\caption{\label{fig:Behavior-of-,} Behavior of $u_{m}$, $\bar{a}$ and $\bar{b}$
in terms of $\zeta$ for Kar's metric. Note that $u_{ms}$ and $\bar{b}_{s}$ are the
values for Schwarzschild metric.}
\end{figure}

\begin{figure}[ht]
\begin{centering}
\includegraphics[scale=0.5]{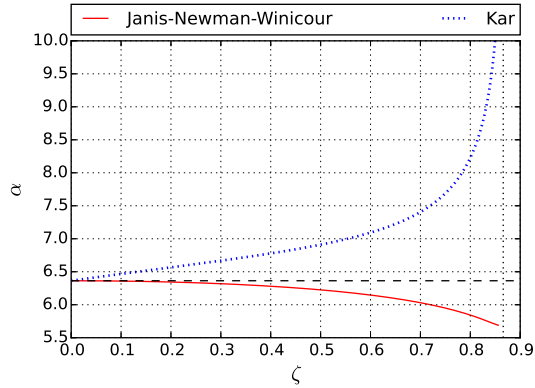}
\par\end{centering}
\protect\caption{\label{fig:Deflection-angles-for-1} Deflection angles for Kar's and JNW's metric
evaluated at $u=u_{m}+0.003$ as a function of $\zeta$. The dashed
horizontal line is the Schwarzschild limit. }
\end{figure}

In Fig. \ref{fig:Deflection-angles-for-1} is plotted the behavior of
the deflection angles as a function of $\zeta$. Once we fix $u=u_{m}+0.003$
we see that the deflection angle has a minimum value at $\zeta=0$, increases as the value of $\zeta$ increases, and for 
$\sqrt{3}/2$ it diverges. For $\zeta=0$
the deflection angle reduces to $-\ln\left(\frac{0.006}{3\sqrt{3}}\right)+\ln(216(7-4\sqrt{3}))-\pi\approx6.364$,
which is the Schwarzschild deflection angle in the strong field limit
when $u=u_{m}+0.003$ \cite{Bozza2002}.

\section{\label{sec:Analysis} Analysis}

It is well known that before the general theory of relativity was proposed, scalar field has been conjectured to give 
rise to the long range gravitational fields. In this sense, several theories involving scalar fields have been 
suggested. One of the most important theories was proposed by C. Brans and R. H. Dicke in \cite{BransDicke}. In that 
paper, the authors discuss the role of the Mach's principle in physics in relation with the equivalence principle, and 
the difficulties to incorporate the former into general theory of relativity. As a consequence, the authors develop a 
new relativistic theory of gravitation (a generalization of general relativity) which is compatible with Mach's 
principle. This theory is not completely geometrical because gravitational effects are described by a scalar field. 
Therefore, the gravitational effects are in part geometrical and in part due to a scalar interaction. According to the 
authors there is a connection between this theory and that of Jordan \cite{Jordan} because the two metric tensor are 
connected by a conformal transformation. Thus, there are two formulations of the Brans-Dicke theory; the so called 
Jordan conformal frame (JF) and the Einstein conformal frame (EF). The equivalence between this two frame of the 
Brans-Dicke (BD) theory of gravity under conformal transformations
\begin{equation}
g_{\mu\nu}^{J}=e^{2\phi}g_{\mu\nu}^{E},\label{conformal trasnformation}
\end{equation}
has been discussed in the literature since long ago \cite{Dicke1962} and, even today, the scientific community is 
divided into two 
points of view: one in which the two frames are equivalent and other in which they are not. In this sense, the question 
of whether or not the two frames are equivalent is very important because there are many applications of scalar-tensor 
theories and of conformal transformation techniques to the physics of early universe and to astrophysics 
\cite{Support}.

The theoretical predictions to be compared with the observations depend on the conformal frame adopted 
\cite{Faraoni} but the discussion is still open. For example, in Ref. \cite{Magnano} is argued that the JF is unphysical, while the EF is physical and 
reveals the physical contents of the theory, in Ref. \cite{Faraoni} 
the authors discussed the 
question of 
which conformal frame is physical by providing a example based on gravitational waves and they favoured the EF as the physical one, in Refs. \cite{Corda} the author showed that the motion of test masses in the field of a 
scalar gravitational wave is different in the two frames and in Ref. \cite{Quiros} is explained an 
alternative interpretation of the conformal transformations of the metric according to which the latter can be viewed 
as 
a mapping among Riemannian and Weyl-integrable spaces, suggesting that these transformations relate complementary 
geometrical pictures of a same physical reality and the question about which is the physical conformal frame, 
does 
not arise. In this sense, the problem of whether the two formulations of a 
scalar-tensor theory in the two conformal frames are equivalent or not is not yet settled, and often is the source of 
confusion in the technical literature \cite{Faraoni,Corda}. 

\begin{figure}[ht]
\begin{centering}
\includegraphics[scale=0.5]{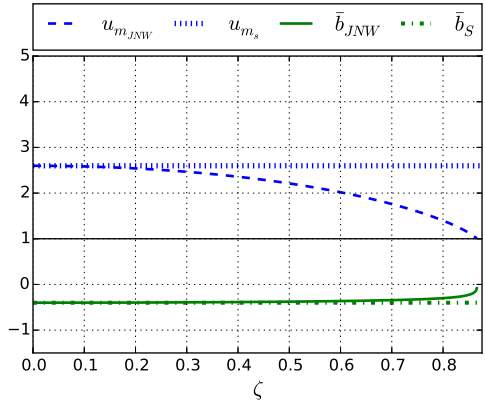}
\par\end{centering}

\protect\caption{\label{fig:Behavior-of-the-1} Behavior of the expansion parameters.
$u_{m}$, $\bar{a}$ and $\bar{b}$ for
JNW's metric in terms of $\zeta$. The parameters with a 's' subscript are
the values for Schwarzschild. }
\end{figure}

Here, in order to compare both frames, we have also calculated the JNW's coefficients and deflection angle (see Eqns. (\ref{correction}) and (\ref{correctionI})) in terms of $\zeta$ as shown in 
Figs. \ref{fig:Behavior-of-the-1} and \ref{fig:Deflection-angles-for-1}. In Fig. \ref{fig:Behavior-of-the-1} we 
plot $\overline{a}$, $\overline{b}$, $u_m$ for JNW's metric as a function of $\zeta$ by making 
$\gamma=\sqrt{1-\zeta^{2}}$. It can be seen that the behavior of $u_{m}$ and $\bar{a}$ are the same in each frame, when 
we compare with Fig. \ref{fig:Behavior-of-,}. Nevertheless, the behavior of $\bar{b}$ is quite 
different for each frame (see once again Fig. \ref{fig:Behavior-of-,}). While $\bar{b}$ for Kar has a minimum at 
$\zeta=0$, $\bar{b}$ for JNW' metric has a minimum near 0.2. This difference
between Kar and JNW's metrics may arise because equations for $\bar{b}$
in each metric are not smooth at $\zeta=\frac{\sqrt{3}}{2}$ (see
equation (\ref{a_and_bdefletion theta})) and $\gamma=0.5$ respectively
in agreement with the ideas presented in \cite{AlvarezConde2002}. On the other hand, the photon sphere equation in terms 
of $\zeta$ for Kar's metric is  (\ref{photon sphere kar}) 
\begin{equation}
x_{m}=\sqrt{1-\zeta^{2}}+\frac{1}{2}.
\end{equation}
At a first glance Eqns. (\ref{JNWxm}) and (\ref{photon sphere kar})
seems to be different; however it is possible to obtain one from the
other by using $\gamma^{2}+\zeta^{2}=1$. Note that for $\zeta=0$ ($\gamma=1$) both reduce to $x_{m}=\frac{3}{2}$,
which is the photon sphere for Schwarzschild, and for $\zeta=\frac{\sqrt{3}}{2}$
($\gamma=\frac{1}{2}$) both diverge. 

Finally in Fig. \ref{fig:Deflection-angles-for-1}, we plot the
deflection angle for JNW's and Kar's metrics.
Although, both frames reduce to Schwarzschild for $\zeta=0$ and
diverge for $\zeta=\frac{\sqrt{3}}{2}$, the behavior of the
deflection angle is very different in each frame.



\subsection{Magnification}

In Ref. \cite{Virbhadra2009} is defined that the magnification of an image formed due to Gravitational lens (GL) is the 
ratio  of the flux of the image to the flux of the unlensed source. Nevertheless, according to Liouville’s theorem, the  
surface  brightness is preserved in GL. Therefore, the magnification $\mu$ of an image turns out to be the ratio of the 
solid  angles of the image and of the unlensed source made at the observer. In this sense, for a circularly symmetric 
gravitational lensing, the magnification of the images can be expressed as \cite{Virbhadra2009} 

\begin{equation}
\mu=\mu_r\mu_t,
\end{equation}
where
\begin{equation}
\begin{aligned}
\mu_t&=&\left(\frac{\sin\delta}{\sin\theta}\right)^{-1}&&\mu_r&=\left(\frac{d\delta}{d\theta}\right)^{-1},
\end{aligned}
\end{equation} 
where $\mu_r$ and $\mu_t$ are the radial and tangential magnification, respectively. 

Tangential critical curves (TCCs) 
and radial critical curves (RCCs) are, respectively, given by singularities in $\mu_t$ and $\mu_r$ in the lens plane. 
However, their corresponding values in the source plane are, respectively,  termed tangential caustic (TC) and radial 
caustics (RCs). The parity of an image is called positive if $\mu> 0$ and negative if $\mu< 0$. If the angular source 
position $\delta=0$ (i.e., when the source, the lens, and the observer are aligned), there may be ring shaped image(s), 
which are called Einstein ring(s) \cite{Virbhadra2009}. The tangential 
magnification is defined as \cite{Bozza2002}

\begin{equation}
\mu^t_n=\left(\frac{\delta}{\theta^0_n}\right)^{-1}\label{tangencial magnification},
\end{equation} 
where 
\begin{equation}
\theta^0_n=\frac{u_m}{D_{OL}}\left(1+e^{\bar{b}-2n\pi}\right).\label{theta_n}
\end{equation}
To study the behavior of the magnification as a function of $\delta$ (the source's position), and compare the behavior 
for Kar's and JNW's metric, we have chosen for this analysis, as the lens, the galactic ``black hole'' with mass 
$M=2.8\times 10^6 M_\odot$ and $D_{OL}=85kpc$ so that $\frac{M}{D_{OL}}\approx 1.75\times10^{-11}$ (geometrized units 
Cfr. \cite{Virbhadra2009}). Therefore, using Eqns. (\ref{b_R para Kar}), (\ref{theta_n}) and (\ref{u_m Kar en fucion 
de zeta}) we obtain that
\begin{equation}
\mu^t_n=\frac{\frac{3\sqrt{3}}{2}}{6.37\times10^{10} 
\delta}(1+e^{2\ln(6(2-\sqrt{3}))-0.226\zeta-\pi+\ln\left[\frac{2\beta_{m}}{y_{m}}\right]-2n\pi}).\label{tangential 
magnification Kar }
\end{equation}
In a similar way, the tangential magnification for JNW's metric, using Eqns. (\ref{correctionI}) and (\ref{tangencial magnification}), is 
\begin{equation}
\mu^t_n=\frac{\frac{3\sqrt{3}}{2}}{6.37\times10^{10} 
\delta}(1+e^{2\ln[6(2-\sqrt{3})]-0.1199(\gamma-1)-\pi+\ln\left(\frac{(2\gamma+1)[(2\gamma+1)^\gamma-(2\gamma-1)^\gamma]
^2,}{2\gamma^2(2\gamma-1)^{2\gamma-1}}\right)-2n\pi}).\label{tangential magnification JNW }
\end{equation}

\begin{figure}[ht]
\begin{centering}
\includegraphics[scale=0.5]{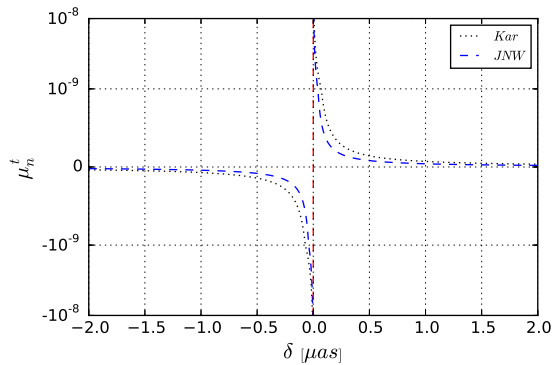}
\par\end{centering}

\protect\caption{\label{fig:Magnification} Tangential magnification $\mu^t_n$ for $\zeta\approx \sqrt{3}/2$ and $n=1$ 
in terms
of $\delta$ for Kar's and JNW's metrics. The lens is the Galactic ‘‘black hole’’. See text for more details. 
}
\end{figure}

\section{\label{sec:Discussion} Discussion}

According to the authors in Ref. \cite{AlvarezConde2002}  ``\textit{there is 
no doubt of the equivalence of all frames for the description of the gravitational effects of the string theories at a 
basic level, at least when all the functions involved are smooth}''. This final statement opens 
the possibility to compare both frames from the strong lensing point of view, since the deflection angle diverges near 
the photon sphere and makes possible not only to differentiate both frames, but also to contribute, in some way, to the 
discussion about the equivalence between EF and JF.

The controversy on conformal frames could appear a purely technical one \cite{Corda}. However, in this paper 
we have shown 
explicitly, at least at first order, that there is a difference in the deflection angle between Jordan and Einstein 
frames, even when some parameters, i.e., $\overline{a}$, $u_m$ and $x_m$, of the strong field expansion in each frame are the same. Nevertheless, considering 
the results presented in Ref. \cite{AlvarezConde2002}, the discrepancies arise due to the strong field expansion, 
which is not smooth near the photon sphere. Our results support the idea that direct evidence from 
observations, in this case from gravitational lensing, could distinguish, if there is, a physical difference between 
the frames. 
However, it is definitively challenging to reach the required relativistic images, see Fig. 
\ref{fig:Magnification}, to discriminate the frames in the 
near future, even with new second-generation Very Large Telescope Interferometer (VLTI), as GRAVITY \cite{GRAVITY} for instance.

\section*{Acknowledgements}

Thanks are due to K. S Virbhadra and F. A. Diaz for valuable discussions and helpful
correspondence, and to the anonymous referees for their constructive inputs. This research was supported by the 
National Astronomical
Observatory, National University of Colombia and one of us, A.C.,
also thanks the Department of Mathematics, Konrad Lorenz University
for financial support. 

\appendix

\section{\label{sec:Appendix}Appendix: Finding $\hat{\alpha}(\theta)$ for Kar's metric} In order to calculate the 
deflection angle in the 
strong field limit, we used Kar's metric in the form proposed in equation (\ref{Kar metric II}), i.e.,

\begin{equation}
ds_{str}^{2}=\underbrace{\left(1-\frac{1}{x}\right)^{\zeta+\sqrt{1-\zeta^{2}}}}_{A(x)}dt^{2}-\underbrace{\left(1-\frac{1
}{x}\right)^{\zeta-\sqrt{1-\zeta^{2}}}}_{B(x)}dx^{2}-\underbrace{\left(1-\frac{1}{x}\right)^{1+\zeta-\sqrt{1-\zeta^{2}}}
x^{2}}_{C(x)}d\Omega^{2}. \label{equation31}
\end{equation}

The deflection angle is calculated using the strong field expansion \cite{Virbhadra2002,Bozza2002}

\begin{equation}
\widehat{\alpha}(\theta)=\overline{a}\ln\left(\frac{\theta D_{OL}}{u_m}-1\right)+\overline{b},
\end{equation}
where
\begin{eqnarray}
\bar{a} & = & \frac{R(0,x_{m})}{2\sqrt{\beta_{m}}},\nonumber \\
\bar{b} & = & b_{R}+\bar{a}\ln\left[\frac{2\beta_{m}}{y_{m}}\right]-\pi
\end{eqnarray}
and
\begin{equation}
u_m=\sqrt{\frac{C(x_{m})}{A(x_{m})}}.
\end{equation}

\subsection{\textbf{Calculation of $\bar{a}$:}}
Using Eqns. (\ref{R(z,x)}), (\ref{f(x,z)}), (\ref{beta en terminos de la metrica}), (\ref{photon sphere kar}) and 
(\ref{equation31}) we obtain that

\begin{equation}
R(z,x_m)=\frac{2\sqrt{1-\zeta^2}+1}{\sqrt{1-\zeta^2}+\zeta}\left\{\frac{\left(\frac{2\sqrt{1-\zeta^2}-1}{2\sqrt{
1-\zeta^2}+1}\right)^\frac{1+\zeta-\sqrt{1-\zeta^2}}{2}-\left(\frac{2\sqrt{1-\zeta^2}-1}{2\sqrt{1-\zeta^2}+1}
\right)^\frac{3\zeta-\sqrt{1-\zeta^2}+1}{2}}{[(1-y_m)z+y_m]^\frac{\zeta}{\sqrt{1-\zeta^2}+\zeta}}\right\}. 
\label{R(z,x_m)}
\end{equation}
Evaluating the last expression at $z=0$ we obtain 

\begin{equation}
R(0,x_m)=\frac{2\sqrt{1-\zeta^2}+1}{\sqrt{1-\zeta^2}+\zeta}\left\{\frac{\left(\frac{2\sqrt{1-\zeta^2}-1}{2\sqrt{
1-\zeta^2}+1}\right)^\frac{1+\zeta-\sqrt{1-\zeta^2}}{2}-\left(\frac{2\sqrt{1-\zeta^2}-1}{2\sqrt{1-\zeta^2}+1}
\right)^\frac{3\zeta-\sqrt{1-\zeta^2}+1}{2}}{y_m^\frac{\zeta}{\sqrt{1-\zeta^2}+\zeta}}\right\} \label{R(0,x_m)}
\end{equation}
where 
\begin{equation}
y_m=\left(\frac{2\sqrt{1-\zeta}-1}{2\sqrt{1-\zeta^2}+1}\right)^{\sqrt{1-\zeta^2}+\zeta}.\label{y_m for Kar}
\end{equation}
From equation (\ref{beta en terminos de la metrica}) we obtain that 
\begin{equation}
\beta_{m}=\frac{1}{4}\frac{\left[\left(2\sqrt{1-\zeta^{2}}+1\right)^{\zeta+\sqrt{1-\zeta^{2}}}-\left(2\sqrt{1-\zeta^{2}}
-1\right)^{\zeta+\sqrt{1-\zeta^{2}}}\right]^{2}}{(\sqrt{1-\zeta^{2}}+\zeta)^{2}(3-4\zeta^{2})^{\sqrt{1-\zeta^{2}}
+\zeta-1}}
\end{equation}
and taking into account equation (\ref{photon sphere kar}) it is possible to express $R(0,x_m)$ and $\beta_m$ and terms 
of the photon sphere as,
\begin{equation}
R(0,x_m)=\frac{2x_m}{k}\left(\left(1-\frac{1}{x_m}\right)^{\frac{1-k}{2}}-\left(1-\frac{1}{x_m}\right)^{\frac{k+1}{2}}
\right).
\end{equation}
 
\begin{equation}
\beta_m=\frac{(x^k_m-(x_m-1)^k)^2}{k^2(x_m-1)^{k-1}x^{k-1}_m}
\end{equation}
then, from equation (\ref{a_and_bdefletion theta})
\begin{equation}
\bar{a}=\frac{x_m[x^{k-1}_m-(x_m-1)^kx^{-1}_m]}{x^k_m-(x_{m}-1)^k}=1.
\end{equation}
 
\subsection{\textbf{Calculation of $u_m$:}}
From equation (\ref{u_m}) the impact parameter is calculated as,
\begin{equation}
u_m=\sqrt{\frac{C(x_m)}{A(x_m)}},
\end{equation}
then, using equation (\ref{equation31}), (\ref{photon sphere kar}) and taking into account that $A(x_m)=y_m$
\begin{equation}
u_m=\frac{(x_m-1)^{\frac{1}{2}+\sqrt{1-\zeta^2}}}{x_m^{-\frac{1}{2}-\sqrt{1-\zeta^2}}}=\frac{1}{2}\frac{(2\sqrt{1-\zeta^
{2}}-1)^{\frac{1}{2}-\sqrt{1-\zeta^{2}}}}{(2\sqrt{1-\zeta^{2}}+1)^{-\frac{1}{2}-\sqrt{1-\zeta^{2}}}}.\label{u_m Kar en 
fucion de zeta}
\end{equation}
\subsection{\textbf{Calculation of $b_R$:}} 
The term $b_R$ is calculated by equation (\ref{b_R}), which corresponds to the regular 
part of the integral (\ref{ID_IR}), i. e.,

\begin{equation}
b_R=I_R(x_m)=\int^1_0g(z,x_m)dz=\int^1_0[R(z,x_m)f(z,x_m)-R(0,x_m)f_0(z,x_m)]dz.
\end{equation}

However, as is pointed out in Ref. \cite{Bozza2002}, $b_R$ can not be calculated analytically but by an expansion of 
$I_R(x_m)$ in powers of some metric's parameter. In this sense, to calculate the regular term for Kar's metric 
(\ref{Kar metric II}), we made 
and expansion of $I_R(x_m)$ in powers of $\zeta$ around $\zeta=0$\footnote{For $\zeta=0$ all the expressions reduce to 
those of Schwarzschild.} and used the first order expansion as our $b_R$. Mathematically this idea is 
 
\begin{equation}
b_R=\sum^\infty_{n=0}\frac{1}{n!}\frac{d^{(n)}I_R(x_m)}{d\zeta^n}(\zeta-0)^n
\end{equation}
and taking terms up to first order in $\zeta$ we have that\footnote{The simbol 
$\left[\frac{dI_R(x_m)}{d\zeta}\right]_{\zeta=0}$ means the derivative evaluated at $\zeta=0$.}
\begin{equation}
I_R(x_m)=I_R(x_m)_{\zeta=0}+\left[\frac{dI_R(x_m)}{d\zeta}\right]_{\zeta=0}\zeta. \label{expansion IR(x_m)}
\end{equation}
In this sense, in order to calculate the regular term $b_R$, it is necessary to find the functional form of $R(z,x_m)$, 
$f(z,x_m)$, $R(0,x_m)$ and $f_0(z,x_m)$. The form of $R(z,x_m)$ and $R(0,x_m)$ are already shown in Eqns. 
(\ref{R(z,x_m)}) and (\ref{R(0,x_m)}). From Eqns. (\ref{f(x,z)}), (\ref{f0}) and (\ref{beta_m_S_kar}) we have that
\begin{equation}
f(z,x_m)=\frac{1}{\sqrt{y_m-[(1-y_m)z+y_m]\frac{C_m}{C}}}\\
\end{equation}
and 
\begin{equation}
f_0(z,x_m)=\frac{1}{\sqrt{\beta_m}|z|}=\frac{2(\sqrt{1-\zeta^2}+\zeta)(3-4\zeta^2)^\frac{\sqrt{1-\zeta^2}+\zeta-1}{2}}{
(2\sqrt{1-\zeta^2}+1)^{\sqrt{1-\zeta}+\zeta}-(2\sqrt{1-\zeta^2}-1)^{\sqrt{1-\zeta}+\zeta}}\frac{1}{|z|},
\end{equation}
where
\begin{equation}
C_m=\left[\frac{2\sqrt{1-\zeta^2}+1}{2}\right]^2\left[\frac{2\sqrt{1-\zeta^2}-1}{2\sqrt{1-\zeta^2}+1}\right]^{
1+\zeta-\sqrt{1-\zeta^2}},
\end{equation}

\begin{equation}
C=\frac{[(1-y_m)z+y_m]^\frac{1+\zeta-\sqrt{1-\zeta^2}}{\zeta+\sqrt{1-\zeta^2}}}{[1-[(1-y_m)z+y_m]^\frac{1}{\zeta+\sqrt{
1-\zeta^2}}]^2}.\\
\end{equation}
\noindent
As equation (\ref{expansion IR(x_m)}) shows, the value of $I_R(x_m)$ reduces to that of Schwarzschild for $\zeta=0$. 
Therefore, for $0\leq z\leq1$ ($|z|=z$) we obtain

\begin{equation}
I_R(x_m)_{\zeta=0}=2\int^1_0\left[\frac{1}{|z|\sqrt{1-\frac{3}{2}z}}-\frac{1}{|z|}\right]dz=2\ln{6(2-\sqrt{3})}=0.9496.
\end{equation}
On the other hand,
\begin{align*}
\left[\frac{dI_R(x_m)}{d\zeta}\right]_{\zeta=0}&=\int^1_0\left[\frac{d}{d\zeta}(R(z,x_m)f(z,x_m))-\frac{d}{d\zeta}(R(0,
x_m)f_0(z,x_m))\right]_{\zeta=0}dz\\
&=\int^1_0[f(z,x_m)\frac{d}{d\zeta}R(z,x_m)
+R(z,x_m)\frac{d}{d\zeta}f(z,x_m)\\
&-f_0(z,x_m)\frac{d}{d\zeta}R(0,x_m)-R(0,x_m)\frac{d}{d\zeta}f_0(z,x_m)]_{\zeta=0}dz\\
&=\int^1_0[f_S(z,x_m)\left[\frac{d}{d\zeta}R(z,x_m)\right]_{\zeta=0}
+2\left[\frac{d}{d\zeta}f(z,x_m)\right]_{\zeta=0}\\
&-f_{0S}(z,x_m)\frac{d}{d\zeta}R(0,x_m)-2\left[\frac{d}{d\zeta}f_0(z,x_m)\right]]dz,
\end{align*}
where $f_S(z,x_m)$, $f_{0S}(z,x_m)$ are those of Schwarzschild (Cfr. \cite{Bozza2002}). Therefore,
\begin{align*}
\left[\frac{dI_R(x_m)}{d\zeta}\right]_{\zeta=0}&=\int^1_0[\frac{\left[\frac{d}{d\zeta}R(z,x_m)\right]_{\zeta
=0}}{z\sqrt{1-\frac{2}{3}z}}-\frac{\left[\frac{d}{d\zeta}R(0,x_m)\right]_{\zeta=0}}{z}\\
&+2\left[\frac{d}{d\zeta}f(z,x_m)\right]_{\zeta=0}-2\frac{\left[\frac{d}{d\zeta}f_0(z,x_m)\right]_{\zeta=0}}{z}]dz,
\end{align*}
where all derivatives, evaluated at $\zeta=0$, are:
\begin{equation}
\begin{aligned}
\left[\frac{d}{d\zeta}R(z,x_m)\right]_{\zeta=0}&=-2-2\ln\left(\frac{2}{3}z+\frac{1}{3}\right)\\
\left[\frac{d}{d\zeta}R(0,x_m)\right]_{\zeta=0}&=-2+2\ln(3)\\
\left[\frac{d}{d\zeta}f_0(z,x_m)\right]_{\zeta=0}&=\frac{\ln(3)-1}{|z|}\\
\left[\frac{d}{d\zeta}f(z,x_m)\right]_{\zeta=0}&=-\frac{1}{2}\frac{\ln{(3)}\left[\frac{7}{3}z^3-2z^2\right]
+z(2z+1)(1-z)\ln(2z+1)}{z^3(1-\frac{2}{3}z)^\frac{3}{2}}.
\end{aligned}
\end{equation}
Thus,

\begin{equation}
\begin{aligned}
\left[\frac{dI_R(x_m)}{d\zeta}\right]_{\zeta=0}&=\int^1_0\left[\frac{2-2\ln(3)}{z}-\frac{2+2\ln\left(\frac{2}{3}z+\frac{
1}{3}\right)}{z\sqrt{1-\frac{2}{3}z}}\right]dz\\
&+\int^1_0\left[\frac{\ln{(3)}\left[\frac{7}{3}z^3-2z^2\right]+z(2z+1)(1-z)\ln(2z+1)}{z^3(1-\frac{2}{3}z)^\frac{3}{2}}
+\frac{2\ln(3)-2}{z}\right]dz\\
&=\int^1_0\left[\frac{2-2\ln(3)}{z}-\frac{2+2\ln\left(\frac{2}{3}z+\frac{1}{3}\right)}{z\sqrt{1-\frac{2}{3}z}}\right]
dz\\
&+\underbrace{\int^1_0\left[\frac{-2\ln(3)z^2+z(2z+1)(1-z)\ln(2z+1)}{z^3(1-\frac{2}{3}z)^\frac{3}{2}}+\frac{2\ln(3)-2}{z
}\right]dz}_i\\
&+\frac{7\ln(3)}{3}\int^1_0\frac{dz}{\left(1-\frac{2}{3}z\right)^{\frac{3}{2}}}.\\
\end{aligned}
\end{equation}

The latter integrals were calculated numerically as

\begin{equation}
\label{la derivada de I_R respecto a zeta}
\begin{aligned}
\int^1_0\left[\frac{-2\ln(3)z^2+z(2z+1)(1-z)\ln(2z+1)}{z^3(1-\frac{2}{3}z)^\frac{3}{2}}+\frac{2\ln(3)-2}{z}\right]
dz&=-2.3980,\\
\int^1_0\left[\frac{2-2\ln(3)}{z}-\frac{2+2\ln\left(\frac{2}{3}z+\frac{1}{3}\right)}{z\sqrt{1-\frac{2}{3}z}}\right]
dz&=-3.457723875,\\
\frac{7\ln(3)}{3}\int^1_0\frac{dz}{\left(1-\frac{2}{3}z\right)^\frac{3}{2}}&=\frac{14\sqrt{3}\ln(3)}{(3+\sqrt{3})}.
\end{aligned}
\end{equation}

Therefore, $b_R$ for Kar's metric is

\begin{equation}
b_R=2\ln(6(2-\sqrt{3}))-0.226\zeta.
\end{equation}
      
Finally, the deflection angle for Kar's metric is 

\begin{equation}
\hat{\alpha}=-\ln\left[\frac{\theta 
D_{OL}}{u_{m}}-1\right]+2\ln(6(2-\sqrt{3}))-0.226\zeta-\pi+\ln\left[\frac{2\beta_{m}}{y_{m}}\right],\label{alphaappend}
\end{equation}
where $u_m$, $\beta_m$ and $y_m$ are given by Eqns. (\ref{u_m Kar en fucion de zeta}), (\ref{beta_m_S_kar}) and 
(\ref{y_m for Kar}) respectively.

%
%




\end{document}